\documentclass[onecolumn,floats]{revtex4}
\usepackage{epsfig}
\usepackage{amssymb,latexsym,amsmath}
\newcommand{\eq}[1]{\begin{align} #1 \end{align}}

\begin{document}

\title
{Quantum Gases in the Microcanonical Ensemble near the Thermodynamic Limit}

\author{V.V. Begun$^1$, M.I. Gorenstein$^{1,2,3}$, A.P. Kostyuk$^{1,2,3}$,
and O.S. Zozulya$^{1,4}$
}
\affiliation{
$^1$ Bogolyubov Institute for Theoretical Physics, Kyiv, Ukraine\\
$^2$ Frankfurt Institute for Advanced Studies, Frankfurt, Germany\\
$^3$ Institut f\"ur Theoretische Physik, Johann Wolfgang Goethe Universit\"at,
 Frankfurt, Germany\\
$^4$ Utrecht University, Utrecht, The Netherlands
}

\begin{abstract}
A new method is proposed for a treatment of ideal quantum
gases in the microcanonical ensemble near the thermodynamic limit.
The method allows rigorous asymptotic calculations of the
average number of particles and particle number fluctuations in
the microcanonical ensemble. It gives also the finite-volume
corrections due to exact energy conservation for the total
average number of particles and for higher moments of the particle 
number distribution in a system
approaching the thermodynamic limit. A present consideration
confirms our previous findings that the scaled variance for
particle number fluctuations in the microcanonical ensemble is
different from that in the grand canonical ensemble even in the
thermodynamic limit.
%
\end{abstract}

\maketitle

\section { Introduction}

 The statistical hadron gas
approach to nucleus-nucleus (A+A) collisions (see e.g.
Ref.~\cite{stat-model} and recent review \cite{PBM}) is rather
successful in describing the data on particle multiplicities in a
wide range of the collision energies. Usually one considers a
thermal system created in A+A collision in the grand canonical
ensemble (GCE). The canonical ensemble (CE)
\cite{ce-a,ce,ce-b,ce-c,ce-d,ce-e} or even the microcanonical
ensemble (MCE) \cite{mce} has been used in order to describe
the $pp$, $p\bar{p}$ and $e^+e^-$ collisions when a small number
of secondary particles are produced.
 In all these cases, the
statistical systems are far away from the thermodynamic limit, so
the statistical ensembles are not equivalent and exact charge
conservation or both energy and charge conservation laws have to
be taken into account. The CE is relevant also for systems with a large
number of produced particles, e.g., a large number of resultant
pions or large nucleon number in p+A collisions, but a small
(of the order of 1 or smaller) number of carriers of conserved
charges, such as strange hadrons \cite{ce-c}, antibaryons
\cite{ce-d} or charmed hadrons \cite{ce-e}. This may happen not
only in elementary  but also in p+A
or even A+A collisions.

The analysis of the fluctuations is a useful tool to study the
properties of the system created during  high energy  particle and
nuclear collisions (see,e.g., Refs.
\cite{GandM,Stod,Shur,Mrow,steph1,steph2,steph3,fluc1} and
 the review papers \cite{fluc}).
In A+A collisions one prefers to use the GCE because it is the
most convenient one from the technical point of view and due to
the fact that both the CE and MCE are equivalent to the GCE in the
thermodynamic limit when the size of the system tends to infinity.
However, the thermodynamic equivalence of ensembles means only
that the {\it average values} of physical quantities calculated in
different ensembles are equal to each other in the thermodynamic
limit.
It was demonstrated for the first time
 in Ref.~\cite{ce-fluc} that {\it multiplicity
fluctuations} are different in the CE
and GCE even in the thermodynamic limit. These results have been then
verified and extended in Refs.~\cite{ce-fluc1,ce-fluc2,ce-fluc3}.
 The particle number
fluctuations in the MCE  have been considered in our paper
\cite{mce-fluc} and they have been shown to be different from
the GCE results even in the thermodynamic limit.

In this paper we present a new method for the study of particle number
distribution in the microcanonical ensemble. The method is based 
on the analysis of the asymptotic behaviour of moments of the particle
number distribution. In contrast to the previously
used microscopic correlator approach \cite{steph2}, the new method
is more rigorous, consistent and mathematically justified. It
elucidates some subtleties. Along with fluctuations, it allows to
calculate the finite-volume corrections to the thermodynamic
quantities.

 The paper is organized as follows. In Section II we review
 the method of microscopic correlator
 and demonstrate its limitations. A detailed description of our
 new method is given in  Section III.  We summarize
 our consideration in Section IV.

\section{Microscopic correlator}

 Let consider the quantum system of non-interacting neutral
 particles. We review here the application of the method
 of Ref.~\cite{steph2} to the calculation of the particle
 number fluctuations in the systems with the exact conservation
 of energy imposed.

The GCE partition function for a single quantum state
with momentum $p$ has the form
\begin{equation}\label{zp}
z_{p} = \sum_{n} \exp\left(- \frac{\epsilon_{p}}{T}~ n\right)~,
\end{equation}
where $T$ is the system temperature, $\epsilon_{p}\equiv \sqrt{p^2+m^2}$
and $m$ is the particle mass. The sum in Eq.~(\ref{zp}) runs over the
number of particles $n=0,1$ for the Fermi statistics and $n=0,1,2,\dots
\infty$ for the Bose statistics. Summing up the two terms for the Fermi
statistics, or the infinite geometric series for the Bose statistics, one
gets:
\begin{equation}\label{zp1}
z_{p} = \left[1~-~\gamma \exp\left(-~ \frac{\epsilon_{p}}{T}\right)
\right]^{-\gamma},
\end{equation}
where $\gamma=+1$ and $\gamma=-1$ for Bose and Fermi statistics,
respectively. The GCE average values are calculated as ($k=1,2$):
\begin{equation}\label{nk}
\langle n^k_p \rangle_{g.c.e.}~ = ~\sum_{n} n^k~ w_{p}(n),
\end{equation}
where
\begin{equation}\label{wp}
w_{p}(n)~=~z_{p}^{-1}~\exp\left(-~ \frac{\epsilon_{p}}{T}~ n\right)
\end{equation}
is the probability to observe $n$ particles in the given quantum state.
%
%
It is easy to see that
\begin{equation}
\langle n^k_p \rangle_{g.c.e.} ~=~ \frac{(-T)^k}{z_p}\, \frac{\partial^k
z_{p}}{\partial \epsilon_{p}^k}~.
\end{equation}
For $k=1$ we get the familiar Fermi (Bose) distribution
\begin{equation}\label{np-aver}
\langle n_p \rangle_{g.c.e.} ~=~
\frac{1}{\exp\left(\frac{\epsilon_{p}}{T}\right)~-~ \gamma}~,
\end{equation}
and for $k=2$
\begin{equation}\label{np2}
\langle n^2_p \rangle_{g.c.e.}~ =~
\frac{\exp\left(\frac{\epsilon_{p}}{T}\right)~+~1}{\left[
\exp\left(\frac{\epsilon_{p}}{T}\right)~-~ \gamma \right]^2}~ =~ \langle
n_p \rangle_{g.c.e.} \left[ 1 + (1+\gamma) \langle n_p \rangle_{g.c.e.}
\right]~.
\end{equation}
From Eqs.~(\ref{np-aver}-\ref{np2}) it follows:
\begin{equation}\label{np-fluc}
\langle (\Delta n_p)^2 \rangle_{g.c.e.}~\equiv~\langle
n_{p}^{2}\rangle_{g.c.e.}~-~ \langle n_{p}\rangle_{g.c.e.}^{2}~=~ \langle
n_p \rangle_{g.c.e.} \left[ 1 -\gamma \langle n_p \rangle_{g.c.e.}
\right]~ \equiv ~v_p^2~.
\end{equation}
It is easy to see that $\gamma=0$ in Eqs.~(\ref{np-aver}-\ref{np-fluc})
corresponds to the Boltzmann approximation.

\begin{figure}[t!]
\epsfig{file=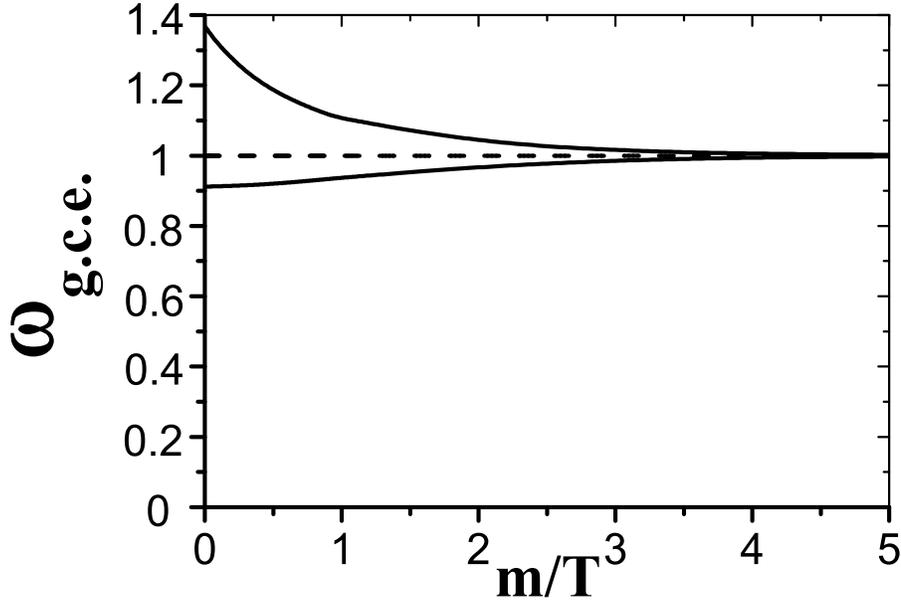,height=8cm,width=12cm}
 \vspace{0cm}
  \caption{The scaled variance (\ref{omega-gce})
  for particle number fluctuation
    in the grand canonical ensemble for different types of statistics.
    The lower and upper solid lines correspond to the Fermi and
    Bose ideal gas, respectively. The dashed line is the Boltzmann
    approximation. }
    \label{fig1}
\end{figure}

Expressions
(\ref{np-aver}) and (\ref{np-fluc}) are microscopic in a sense that they
describe
 the average values and fluctuation of a single mode with momentum $p$.
 However, the fluctuations of macroscopic quantities of the system can be determined
  through the fluctuations of these single modes. To
  be more precise, we will demonstrate that the fluctuations can be written
  in terms of the microscopic correlator $\langle \Delta n_p \Delta
  n_k \rangle_{g.c.e.}$.
  This correlator can be presented as:
  \eq{
  \langle \Delta n_p \Delta n_k \rangle_{g.c.e.}~ =~
  v_p^{2}~
  \delta_{pk}~.  \label{correlator1}
  }
 The variance $\langle(\Delta N)^2\rangle \equiv
 \langle N^{2}\rangle - \langle N \rangle^{2}$
 of the total number of particles,
  $N = \sum_p n_p$, equals to:
 \eq{\langle (\Delta N)^2\rangle_{g.c.e}~
 =~ \sum_{p,k} \langle n_p n_k \rangle_{g.c.e.} -
  \langle n_p \rangle \langle n_k \rangle_{g.c.e.}~
  =~ \sum_{p,k} \langle \Delta n_{p} \Delta n_k
 \rangle_{g.c.e.}~ = ~\sum_p v_p^{ 2}~.
}
 We have assumed above that the quantum $p$-levels
 are non-degenerate. In fact each this level should be
 further specified by the projection of a particle
 spin. Thus each $p$-level splits into $g=2j+1$
 sub-levels. It will be assumed that the $p$-summation includes all
 sub-levels too.
 This does not change the above formulation
 because of statistical independence of these quantum sub-levels.
 The degeneracy factor enters explicitly when one substitutes, in the
 thermodynamic limit,
the summation over discrete levels by the integration:
\eq{\sum_{p}~...~\simeq~\frac{gV}{2\pi^{2}}\int_{0}^{\infty}p^{2}dp~...~.}
The scaled variance $\omega$ in the thermodynamic limit
$V\rightarrow\infty$ reads:
 \eq{\omega_{g.c.e.} ~ \equiv~
 \frac{\langle \left(\Delta N^2\right)\rangle_{g.c.e.}}
 {\langle N \rangle_{g.c.e.}}~
=~ \frac{\sum_{p,k} \langle
 \Delta n_{p} \Delta n_{k} \rangle_{g.c.e.}}
 {\sum_p \langle n_p\rangle_{g.c.e.}}~ = ~
 \frac{\sum_{p}v_p^{ 2}}{\sum_{p}\langle n_p\rangle_{g.c.e.}}~\simeq~
\frac{\int_{0}^{\infty}p^{2}dp~v_p^{ 2}}{\int_{0}^{\infty}p^{2}dp~\langle
n_p\rangle_{g.c.e. } }~. \label{omega-gce}
 }
The Eq.~(\ref{omega-gce}) corresponds to the particle number fluctuations
in the GCE. From Eq.~(\ref{omega-gce}) one finds $\omega_{g.c.e.}=1$ in
the classical Boltzmann limit ($\gamma=0$). The effects of quantum
statistics lead to $\omega_{g.c.e.}>1$ for the Bose gas ($\gamma=1$) and
$\omega_{g.c.e.}<1$ for the Fermi gas ($\gamma=-1$). The strongest effect
corresponds to $m/T\rightarrow 0$,
\eq{
\omega_{g.c.e.}^{Bose}=\pi^{2}/6\zeta(3) ~\simeq
~1.368~,~~~\omega_{g.c.e.}^{Fermi}~=~\pi^{2}/9\zeta(3)~ \simeq~ 0.912~,
\label{omegaBF}
}
and it decreases with increasing $m/T$ (see Fig.~\ref{fig1}).

The formula for the microscopic correlator is modified if we impose the
exact conservation law on our equilibrated system. For this purpose we
introduce the equilibrium probability distribution $W(n_p)$ of the
occupation numbers $\{n_{p}\}$. As a first step we assume that each
$n_{p}$ fluctuates independently according
 to the Gauss distribution law with mean square deviation $v_p^{ 2}$:
 \eq{
   W(n_p) ~\propto ~ \prod_{p}
   \exp{\left[-~ \frac{\left(\Delta n_p\right)^2}{2v_p^{ 2}}
    \right]}~.
\label{gauss} }
To justify this assumption (see Ref.~\cite{steph2}) one can consider
  the
  sum of $n_p$ in small momentum volume $(\Delta  p)^3$
  with the center at $p$. At fixed $(\Delta p)^3$
  and $V\rightarrow \infty$ the average number of particles
  inside $(\Delta p)^3$ becomes large.
  Each particle configuration inside $(\Delta p)^3$
   consists of $(\Delta p)^3 \cdot V/(2\pi)^{3}>>1$ statistically
  independent terms, each
with average value $\langle n_{p}\rangle$ (\ref{np-aver}) and scaled
variance $v^{2}_{p}$ (\ref{np-fluc}). From the central limit
  theorem it then follows that the probability distribution for the fluctuations
  inside $(\Delta p)^3$ should be Gaussian.
In fact, we always convolve $n_p$ with some smooth
  function  of $p$, so instead of writing the Gaussian
  distribution for the sum of $n_p$ in $(\Delta p)^3$
we can use it directly for $n_{p}$.

\begin{figure}[t!]
\epsfig{file=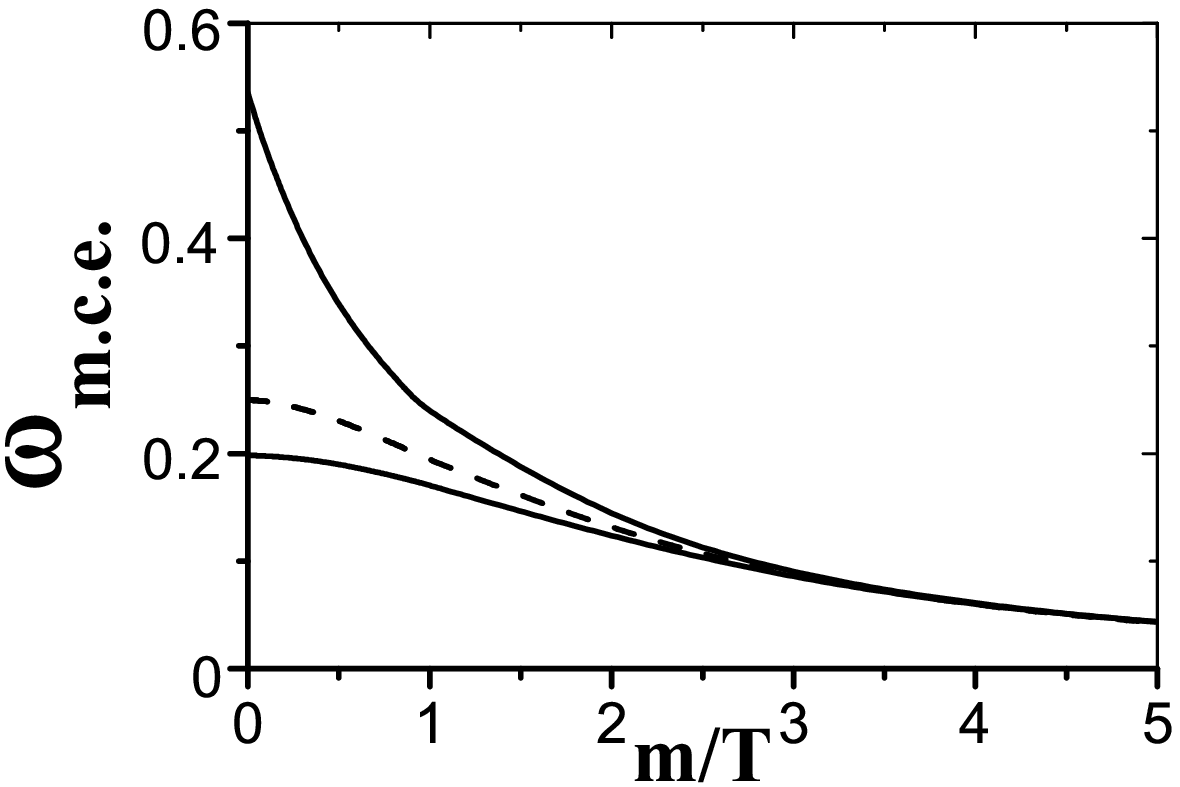,height=8cm,width=12cm}
 \vspace{0cm}
  \caption{The scaled variance (\ref{omega-mce})
  for particle number fluctuation
    in the microcanonical ensemble for different types of statistics.
    The lower and upper solid lines correspond to the Fermi and
    Bose ideal gas, respectively. The dashed line is the Boltzmann
    approximation.}
    \label{fig2}
\end{figure}

Now we want to impose the exact conservation laws. The conserved
quantity $A$ (the energy and/or conserved charge) can be written
in the form $A\equiv\sum_{p}a(p)\,n_{p}$. An exact conservation
law means the restriction on the set of occupation numbers
$\{n_{p}\}$: only those which satisfy the condition $\Delta
A=\sum_{p} a(p)\Delta n_p=0$ can be realized. Let us consider 
exact energy conservation. Then $\;A\rightarrow E\;$ (i.e.
$\;a(p)\rightarrow\epsilon_p\;$) and the distribution
(\ref{gauss}) will be modified because of the energy conservation
as:
\begin{align}\label{gauss-Q}
   W(n_p) ~\propto ~\prod_{p}
   \exp\left[-~ \frac{\left(\Delta n_p\right)^2}{2v_p^{ 2}}
      \right]~
    \delta\left(\sum_{p}\epsilon_p \Delta n_p^{}
    \right)
 \propto ~  \int_{-\infty}^{\infty} d \lambda~\prod_{p}
  \exp\left[-~ \frac{\left(\Delta n_p\right)^2}{2v_p^{ 2}}
+ i \lambda~\epsilon_p \Delta n_p \right]~,
\end{align}
where $\;\delta\left(\epsilon_p \Delta n_p\right)\;$ is the Dirac's
$\delta-$function.
 It is convenient to generalize distribution (\ref{gauss-Q})
 using further the integration along imaginary axis in
 $\lambda$-space.
After completing squares one gets:
   \eq{ W(n_p, \lambda)~ \propto ~  \prod_{p}
    \exp\left[-
    ~\frac{\left(\Delta n_p - \lambda v_p^{ 2}\epsilon_p\right)^2}
    {2v_p^{ 2}}+
    \frac{\lambda^2}{2} v_p^{ 2}\epsilon_p^2\right]~,
    }
and the average values (i.e. the MCE averages) are now calculated as:
\eq{\langle ... \rangle~=~\frac{\int_{-i\infty}^{i\infty}d\lambda
\int_{-\infty}^{\infty}\prod_{p} dn_p^{}~... ~W(n_p,
\lambda)}{\int_{-i\infty}^{i\infty}d\lambda
\int_{-\infty}^{\infty}\prod_{p,} dn_p~W(n_p, \lambda)}~. \label{average}
}
    Using Eq.~(\ref{average}) one easily deduces
    \eq{
    \langle(\Delta n_p~
    v_p^{ 2}\lambda \epsilon_p)(\Delta n_k - v_k^{ 2}\lambda
    \epsilon_k)   \rangle ~=~ \delta_{pk}~ v_p^{ 2}~,
    ~~~
   \langle \lambda^2 \rangle = - \left( \sum_{p}
   v_p^{ 2} \epsilon_p^2 \right)^{-1}~,~~~
  \langle (\Delta n_p^{} - v_p^{ 2}\lambda \epsilon_p) \lambda
  \rangle ~=~ 0~.\nonumber
  }
 Therefore, one finds the MCE average for the microscopic correlator
  \begin{align}\label{mce-corr}
   \langle \Delta n_p \Delta n_k \rangle_{m.c.e.} ~& =~
   \delta_{pk}~ - ~ v_p^{ 2} \epsilon_p~
   v_k^{ 2} \epsilon_k~ \langle \lambda^2 \rangle ~ + ~
   \langle \Delta n_p \lambda \rangle~ v_k^{ 2}
   \epsilon_k + \langle \Delta n_k \lambda \rangle~ v_p^{ 2}
   \epsilon_p \nonumber
   \\
   &=~ \delta_{pk}~  +~ v_p^{ 2} \epsilon_p~
   v_k^{ 2} \epsilon_k ~\langle \lambda^2 \rangle~
   = ~\delta_{pk}~  v_p^{ 2} -
   \frac{v^{ 2}_p \epsilon_p~v^{ 2}_k  \epsilon_k}
   {\sum_{p} v^{ 2}_p \epsilon_p^2}~.
  \end{align}
   By means of Eq.~(\ref{mce-corr}) one obtains:
   \eq{ \label{omega-mce} \omega_{m.c.e.}~&\equiv~\frac{
    \langle(\Delta N^2) \rangle_{m.c.e.} }
    {\langle N \rangle_{m.c.e.}} ~=~\frac{
    \sum_p v_{p}^{ 2}}{\sum_{p}\langle n_{p}\rangle_{g.c.e.}}~-~
    \frac{\left(\sum_p v_{p}^{ 2}\epsilon_p\right)^{2}}
    {\sum_{p}\langle n_{p}\rangle_{g.c.e.}~\sum_{p} v_{p}^{
    2}\epsilon_p^2} \nonumber \\
    ~&\simeq~ \frac{
    \int_{0}^{\infty}p^{2}dp~ v_{p}^{2}}{\int_{0}^{\infty}p^{2}dp~
    \langle n_{p}\rangle_{g.c.e.}}~-~
    \frac{\left(\int_{0}^{\infty}p^{2}dp~ v_{p}^{2}\epsilon_p\right)^{2}}
    {\int_{0}^{\infty}p^{2}dp~\langle n_{p}\rangle_{g.c.e.}~
    \int_{0}^{\infty}p^{2}dp~
     v_{p}^{2}\epsilon_p^2
    } ~.
   }
The Eq.~(\ref{omega-mce}) demonstrates that the MCE fluctuations in the
thermodynamic limit $V\rightarrow\infty$ can be presented in terms of the
GCE quantities. The MCE scaled variances (\ref{omega-mce}) for different
statistics are shown as functions of $m/T$ in Fig.~\ref{fig2}.

The microcanonical suppression effect for the particle number fluctuations
increases with the particle mass. In the Boltzmann approximation
($\gamma=0$, $v_{p}^{2}=\langle
n_{p}\rangle_{g.c.e.}=\exp(-\epsilon_{p}/T)$) the integrals in
Eq.~(\ref{omega-mce}) can be calculated analytically:
\begin{align}
\int_0^{\infty}p^{2}dp~ \exp\left(-\frac{\epsilon_{p}}{T}\right)
  \; &=\; T\,m^2K_2\left(\frac{m}{T}\right)~,
  \\
\int_0^{\infty} p^{2} dp~
\epsilon_{p}~\exp\left(-\frac{\epsilon_{p}}{T}\right)\; &=\;
\frac{m^4}{8}\left[\;K_4\left(\frac{m}{T}\right)~-
~K_0\left(\frac{m}{T}\right)\;\right]~,
  \\
 \int_0^{\infty} p^{2} dp~
\epsilon_{p}~\exp\left(-\frac{\epsilon_{p}}{T}\right)\; &=\;
\frac{m^5}{16}\left[K_5\left(\frac{m}{T}\right)~
    + ~K_3\left(\frac{m}{T}\right)~ -
    ~2\,K_1\left(\frac{m}{T}\right) \;\right]~.
\end{align}
Making use of the asymptotic behavior of the modified Hankel function
$\;K_n(x)\;$ at $x\rightarrow 0$ ($K_0(x) \simeq-\ln x\;$ and
$K_n(x)\simeq
 \frac{1}{2}\;\Gamma(n)\left(\frac{x}{2}\right)^{-n}\;$ for $n\geq 1$)
one gets in the massless limit:
\begin{align}\label{omegaBoltz}
\omega_{m.c.e.}(m=0)~=~\frac{1}{4}~,
\end{align}
i.e. for classical massless particles the MCE the scaled
variance is quarter as large as the corresponding scaled variance
in the GCE. For the case of Bose and Fermi statistic we obtain:
\begin{align}
\omega_{m.c.e.}^{Bose}(m=0)~&=~ \frac{\pi^2}{6\xi(3)}~ -~
\frac{135\xi(3)}{2
\pi^4} ~\simeq ~0.535\;, \label{omegaB}\\
\omega_{m.c.e.}^{Fermi}(m=0)~&= ~\frac{\pi^2}{9\xi(3)}~ - ~\frac{405
\xi(3)}{7 \pi^4} ~\simeq~ 0.198\;. \label{omegaF}
\end{align}
The $\;\omega_{m.c.e.}\;$ for bosons is always larger then in the
Boltzmann limit and for fermions is always smaller, so we acquire the
Bose-enhancement and Fermi-suppression of the fluctuations. They exist
also in the GCE. However, the effects due to the quantum statistics in the
MCE become stronger than those in the GCE (compare
Eqs.~(\ref{omegaBoltz}-\ref{omegaF}) and Eq.~(\ref{omegaBF})).

As it is seen from Fig.~\ref{fig2} the MCE suppression effects for
the fluctuations of massive particles increase with particle mass. 
One finds that with
increasing $m/T$ ratio the scaled variances for the Bose and Fermi
systems approach their Boltzmann limit, $\omega_{m.c.e.}$, and
from the asymptotic behavior, $K_n(x)\simeq
\sqrt{\frac{\pi}{2x}}\,\exp(-x)\left(1+\frac{4n^2-1}{8x}\right)$
at $x\gg 1$, it follows  that
 $\omega_{m.c.e.}\simeq 3/2(m/T)^{-2}\rightarrow 0$
 at $m/T\rightarrow \infty$.

Comparing Eq.~(\ref{mce-corr}) and Eq.~(\ref{correlator1}) one
finds the changes of the microscopic correlator due to the exact
energy conservation. First, in the MCE the fluctuations of each
mode $p$ is reduced, i.e. the $\langle \left(\Delta
n_{p}\right)^{2}\rangle$ calculated from Eq.~(\ref{mce-corr}) is
smaller that calculated from Eq.~(\ref{correlator1}). Second, the
anticorrelations between different modes $p\neq k$ absent in the
GCE (\ref{correlator1}) appear in the MCE (\ref{mce-corr}). These
changes of the microscopic correlator result in the suppression
effect of the MCE scaled variance $\omega_{m.c.e.}$
(\ref{omega-mce}) in a comparison to the GCE one $\omega_{g.c.e}$
(\ref{omega-gce}).

The method considered above has several serious limitations
\begin{itemize}
\item The probability distribution for $\Delta n_p$ was replaced
by the Gaussian distribution (\ref{gauss}), but in fact, this two
distributions are very different. It is enough to mention that the
real distribution is a discrete,  and the Gaussian one is continuous.
Although the arguments given after the equation (\ref{gauss})
suggest that this replacement does not influence the final result
in the thermodynamic limit, still, it would be nice to have a more
rigorous justification of this fact as well as a recipe for the
treatment of finite-size corrections in quantum systems.
\item  The Exact energy conservation influences not only
fluctuations, but also the average values $\langle n_{p}\rangle$
and  $\langle n_{p}^2 \rangle$, and this should result in
microcanonical corrections to $ \langle N\rangle_{m.c.e.}$ and
$\langle N^2 \rangle_{m.c.e.}$ in the case of finite-volume
systems. The approach presented in this section does not allow
calculating these corrections.
\item It would be desirable to convince ourselves that the present
approach gives a correct value for $\omega_{m.c.e.}$ despite of
the fact that it ignores the finite-volume corrections to $
\langle N\rangle_{m.c.e.}^2$ and $\langle N^2 \rangle_{m.c.e.}$
which are of the same order as fluctuations $\langle (\Delta N)^2
\rangle_{m.c.e.}=\langle N^2 \rangle_{m.c.e.} - \langle
N\rangle_{m.c.e.}^2$.
\end{itemize}

The rigorous method described in the next sections resolves the
above issues.


\section{The moments of the particle number distributions
in the MCE}

In this section we consider a consistent and mathematically
justified method for MCE treatment of an ideal gas near and in the
thermodynamical limit. Our aim is to find the asymptotic behavior
of the thermodynamic quantities (the total particle number and its
fluctuations) of a microcanonical thermodynamic system at large
volume $V$ in terms of the thermodynamic properties of a grand
canonical system, which are easier to calculate.

The methods is based on the analysis of the particle number 
distribution and its moments.
For pedagogical purposes, we first apply our approach to the grand
canonical system and than use the gained experience for the
treatment of a microcanonical system near the thermodynamic
limit $V \rightarrow \infty$.

The grand canonical partition function of the ideal quantum gas
is given by a product of the grand canonical partition functions
$z_{p}$ (\ref{zp1}) for each quantum level $p$:
\begin{equation}
Z_{g.c.e.}(T) =  \prod_{p} z_{p} =
\prod_{p} \sum_{n_p}
\exp \left( -\frac{ \epsilon_{p} n_{p}}{T} \right) =
\sum_{\{ n_p \} }
W_{g.c.e.}(\{ n_{p} \})~, \label{Zgce}
\end{equation}
where the sum runs over all possible
sets of the occupation numbers $\{ n_p \}$. The quantity
\begin{equation}
W_{g.c.e.}(\{ n_{p} \}) =
\exp \left( -\frac{\sum_{p} \epsilon_{p} n_{p}}{T} \right) =
\prod_{p} z_{p} w_{p} (n_{p})
\end{equation}
is proportional to the probability to observe a given set $\{
n_{p} \}$ of the occupation numbers. Here $w_{p}(n_{p})$ is the
probability (\ref{wp}) to observe $n_{p}$ particles at the level
$p$.

To get the (unnormalized) probability distribution for the total
particle number we multiply the above expression with the
$\delta$-function $\delta \left( \sum_{p}  n_{p}  - N \right)$ and
sum over all $n_{p}$:
\begin{equation}
W_{g.c.e.}(N) = \sum_{ \{ n_{p} \} } W_{g.c.e.}(\{ n_{p} \}) \,
\delta \left( \sum_{p} n_{p}  - N \right)~.  \label{WgceN}
\end{equation}

Now let us consider the Fourier transform (\ref{Wtx})
of the above probability distribution:
\begin{equation}
\tilde{W}_{g.c.e.}(\nu) = \sum_N \exp \left( i \nu N \right)
W_{g.c.e.}(N) ~.\label{Wtnu}
\end{equation}

The $\delta$-function disappears due to the summation:
\begin{equation}
\sum_N \exp \left( i \nu N \right) \delta \left( \sum_{p} n_{p}  -
N \right) = \exp \left( i \nu  \sum_{p} n_{p} \right)~,
\label{dd}
\end{equation}
this makes the expression (\ref{Wtnu}) factorizable
\begin{eqnarray}
\tilde{W}_{g.c.e.}(\nu) =
\sum_{ \{ n_{p} \} } \exp \left( i \nu  \sum_{p} n_{p} \right)
W_{g.c.e.}(\{ n_{p} \}) =
\prod_{p} z_{p} \sum_{n_p} \exp \left( i \nu  n_{p} \right)
w_{p} (n_{p})
= \prod_{p} \tilde{w}_{p} (\nu) z_{p}~. \label{Wgcenu1}
\end{eqnarray}
Here we have introduced a Fourier transform $\tilde{w}_{p} (\nu)$
of the single-level probability distribution $w_{p}(n_{p})$:
\begin{equation}
\tilde{w}_{p} (\nu) = \sum_{n_p}
\exp \left( i \nu  n_{p} \right) w_{p} (n_{p})~.
\end{equation}
One can rewrite the equation (\ref{Wgcenu1}) as
\begin{equation}
\tilde{W}_{g.c.e.}(\nu) =
\exp \left( \sum_{p} \log \tilde{w}_{p} (\nu)
\right) \prod_{p} z_{p}~ .
\end{equation}
The summation in the exponential can be replaced by
integration, if the system is sufficiently large:
\begin{equation}
\tilde{W}_{g.c.e.}(\nu) =
\exp \left[\frac{g V}{(2 \pi)^3}
\int d^3 \! p \log  \tilde{w}_{p} \left( \nu\right) \right]
  \prod_{p} z_{p} ~.
 \label{Wgcenu2}
\end{equation}

Let us expand the logarithm in the integral into a Taylor series:
\begin{equation}
\log  \tilde{w}_{p} \left( \nu \right) =
\log  \tilde{w}_{p} \left( 0 \right) + \sum_{j=1}^{\infty} i^j
\frac{m_j(p)}{j!} \nu^j~, \label{logwg}
\end{equation}
where, in particular, (see Appendix)
\begin{eqnarray}
m_1(p) &=&  \langle n_p \rangle_{g.c.e.}~, \\
m_2(p) &=& \langle (n_p - \langle n_p \rangle)^2 \rangle_{g.c.e.}
\equiv v_p^2~.
\end{eqnarray}
If one uses the normalized probability distribution for
$w_{p}(n)$, then the first term in (\ref{logwg}) is zero
and the expression (\ref{Wgcenu2}) takes the form
\begin{equation}
\tilde{W}_{g.c.e.}(\nu) =
\exp \left[ \sum_{j=1}^{\infty} i^j
\frac{\nu^j}{j!}
\frac{g V}{(2 \pi)^3}
\int d^3 \! p \, m_j(p)  \right]  \prod_{p} z_{p}~.
 \label{Wgcenu3}
\end{equation}

Now one can calculate the  moments of the
probability distribution $W_{g.c.e.}(N)$ using
the values of the function $\tilde{W}_{g.c.e.}(\nu)$ and
its derivatives at $\nu = 0$ (see Appendix). The value
\begin{equation}
\tilde{W}_{g.c.e.}(0) =  \prod_{p} z_{p}
\end{equation}
is used for the normalization.
The first derivative is related to the average number
of particles:
\begin{equation}\label{Ngce}
\langle N \rangle_{g.c.e.} = \frac{1}{i \tilde{W}_{~g.c.e~}(0)}
\left. \frac{d \tilde{W}_{g.c.e.}(\nu)}{d \nu} \right|_{\nu=0} =
\frac{g V}{(2 \pi)^3} \int d^3 \! p \,
\langle n_p \rangle_{g.c.e.} \equiv \bar{N}~.
\end{equation}
Similarly,
\begin{equation}\label{N2gce}
\langle N^2 \rangle_{g.c.e.} = - \frac{1}{\tilde{W}_{g.c.e.}(0)}
\left. \frac{d^2 \tilde{W}_{g.c.e.}(\nu)}{d \nu^2} \right|_{\nu=0}
= \bar{N}^2 +  \frac{g V}{(2 \pi)^3} \int d^3 \! p \, v_p^2 ~.
\end{equation}

The above results for the GCE are known from the
textbooks. The purpose of this consideration is to
demonstrate how our method works and  to clarify our further step --
a treatment of the MCE.
%
The MCE partition function of the ideal quantum gas is
given by
\begin{equation}
Z_{m.c.e.}(\{n_{p}\}) =  \sum_{ \{ n_{p}\} }
\delta \left( \sum_{p} \epsilon_{p} n_{p} - E \right),
\end{equation}
where the sum, similarly to Eq.~(\ref{Zgce}), runs over all sets of
the occupation numbers $\{n_{p}\}$.
The probability to observe a given set $\{n_{p}\}$ is proportional to
\begin{equation}
W_{m.c.e.}(\{n_{p}\}) =  \delta \left( \sum_{p} \epsilon_{p} n_{p}
- E \right)~.
\end{equation}
Nothing changes if we multiply the last
expression by $1=\exp(E/T) \exp(-E/T)$:
\begin{equation}
W_{m.c.e.}(\{n_{p}\}) = \exp \left( \frac{E}{T} \right) \exp
\left( -\frac{E}{T} \right) \delta \left( \sum_{p} \epsilon_{p}
n_{p} - E \right).
\end{equation}
The parameter $T$ has the meaning of the temperature of a
grand canonical system, which will be defined
later.

Using the properties of the $\delta$-function, one can replace $E$
in the second exponent by $\sum_{p} \epsilon_{p} n_{p}$:
\begin{equation}\label{Wmce}
W_{m.c.e.}(\{n_{p}\}) = \exp \left( \frac{E}{T} \right) \exp
\left( -\frac{\sum_{p} \epsilon_{p} n_{p}}{T} \right) \delta
\left( \sum_{p} \epsilon_{p} n_{p} - E \right)~.
\end{equation}
Then the second exponential function can be rewritten using
Eq.~(\ref{wp}) as
\begin{equation}
\exp \left( -\frac{\sum_{p} \epsilon_{p} n_{p}}{T} \right) =
\prod_{p} \exp \left( -\frac{ \epsilon_{p} n_{p}}{T} \right) =
\prod_{p} w_{p} (n_{p})  z_{p}~,
\end{equation}
where $z_{p}$ is the {\it grand} canonical partition function
(\ref{zp1}) for the single $p$-level, $w_{p}(n_{p})$ (\ref{wp})
is the probability
 to observe $n_{p}$ particles at this level. The grand canonical
system is assumed to have the same quantum levels as the microcanonical
system under consideration.
The $\delta$-function in (\ref{Wmce}) can be represented as
\begin{equation}
\delta \left( \sum_{p} \epsilon_{p} n_{p}  - E \right) =
\frac{1}{2 \pi} \int_{-\infty}^{+\infty} d \lambda \exp \left[ i
\lambda \left( \sum_{p} \epsilon_{p} n_{p}  - E \right) \right]~,
\end{equation}
therefore,
\begin{eqnarray}
W_{m.c.e.}(\{ n_{p} \}) &=& \frac{1}{2 \pi} \exp \left( \frac{E}{T}
\right) \int_{-\infty}^{+\infty} d \lambda \exp \left( - i \lambda
E \right) \prod_{p} w_{p}(n_{p})  z_{p} \exp \left( i \lambda n_{p}
\epsilon_{p} \right) \nonumber \\
&=&
C \int_{-\infty}^{+\infty} d \lambda \exp \left( - i \lambda
E \right) \prod_{p}  w_{p}(n_{p}) \exp \left( i \lambda n_{p}
\epsilon_{p} \right)~,
\end{eqnarray}
where we have introduced the notation
\begin{equation}
C = \frac{1}{2 \pi} \exp \left( \frac{E}{T} \right) \prod_{p} z_{p}~ .
\end{equation}
The constant $C$ is a normalization factor and is irrelevant to the
physical quantities we are interested in.

Similarly to Eq.~(\ref{WgceN}), we get the (unnormalized)
probability distribution for the total
particle number by multiplying the above expression with the
$\delta$-function $\delta \left( \sum_{p}  n_{p}  - N \right)$ and
summing over all possible sets $\{ n_{p} \}$:
\begin{equation}
W_{m.c.e.}(N) = C \int_{-\infty}^{+\infty} d \lambda \exp \left( - i
\lambda E \right) \prod_{p} \sum_{ n_{p} } w_{p} (n_{p}) \exp
\left( i \lambda n_{p} \epsilon_{p} \right) \delta \left( \sum_{p}
n_{p}  - N \right)~.
\end{equation}
We perform a Fourier transformation of $W_{m.c.e.}(N)$:
\begin{equation}
\tilde{W}_{m.c.e.}(\nu) = \sum_N \exp \left( i \nu N \right)~.
W_{m.c.e.}(N)
\end{equation}
The $\delta$-function disappears due to (\ref{dd}), and
the integrand becomes factorizable:
\begin{eqnarray}
\tilde{W}_{m.c.e.}(\nu) &=& C  \int_{-\infty}^{+\infty} d
\lambda \exp \left( - i \lambda E \right) \prod_{p} \sum_{ n_{p} }
w_{p} (n_{p}) \exp \left[ i n_{p} \left( \nu + \lambda
\epsilon_{p} \right) \right] \nonumber \\ &=& C
\int_{-\infty}^{+\infty} d \lambda \exp \left( - i \lambda E
\right) \prod_{p} \tilde{w}_{p} \left( \nu + \lambda \epsilon_{p}
\right) = C \int_{-\infty}^{+\infty} d \lambda
\exp \left[ - i \lambda E + \sum_{p} \log  \tilde{w}_{p} \left(
\nu + \lambda \epsilon_{p} \right) \right]~.
\end{eqnarray}
We replace the summation in the exponential by integration
\begin{equation}
\tilde{W}_{m.c.e.}(\nu) = C \int_{-\infty}^{+\infty}
d \lambda \exp \left[ - i \lambda E + \frac{g V}{(2 \pi)^3}
\int d^3 \! p \log  \tilde{w}_{p} \left( \nu + \lambda
\epsilon_{p} \right) \right]~,
\end{equation}
and  expand the logarithm in the integral into the Taylor
series:
\begin{equation}
\log  \tilde{w}_{p} \left( \nu + \lambda \epsilon_{p} \right) =
\log  \tilde{w}_{p} \left( 0 \right) + \sum_{j=1}^{\infty} i^j
\frac{m_j(p)}{j!} \left( \nu + \lambda \epsilon_{p} \right)^j~.
\end{equation}
This yields
\begin{equation}
\tilde{W}_{m.c.e.}(\nu) = C \int_{-\infty}^{+\infty}
d \lambda \exp \left[ - i \lambda E +
\sum_{j=1}^{\infty}  \frac{i^j}{j!}
\frac{g V}{(2 \pi)^3}
\int d^3 \! p \, m_j(p)
\left( \nu + \lambda \epsilon_{p} \right)^j
\right]~.
\end{equation}
The first term of the sum in the exponential can be rewritten
as
\begin{equation}
i \frac{V}{(2 \pi)^3} \int d^3 \! p \langle n_p \rangle
\left( \nu + \lambda \epsilon_{p} \right) = i \left( \nu \bar{N}
+ \lambda \bar{E} \right)~,
\end{equation}
where $\bar{N}$ is the average total particle number (\ref{Ngce})
in the grand canonical system, and
\begin{equation}
\bar{E} = \frac{V}{(2 \pi)^3} \int d^3 \! p
\epsilon_{p} \langle n_p \rangle
\end{equation}
is its average total energy. The temperature of the GCE has been
arbitrary so far. Now we fix it so that the {\it average} energy
of the GCE equals the energy of the MCE:
$\bar{E}  = E$.
In this case the term $- i \lambda E$ is annihilated by $i \lambda
\bar{E}$. Finally one gets
\begin{equation}\label{Wtildenu}
\tilde{W}_{m.c.e.}(\nu) = C \int_{-\infty}^{+\infty}
d \lambda \exp \left[ i \nu \bar{N} + \sum_{j=2}^{\infty}
\frac{i^j}{j!} \frac{V}{(2 \pi)^3} \int d^3 \! p m_j(p)
\left( \nu + \lambda \epsilon_{p} \right)^j \right]~.
\end{equation}

Let us find the values of the function
$\tilde{W}_{m.c.e.}(\nu)$ and
its derivatives at $\nu = 0$  as they will be needed
to calculate the moments of the
probability distribution $W_{m.c.e.}(N)$. From Eq.~(\ref{Wtildenu})
it follows
\begin{equation}
\tilde{W}_{m.c.e.}(0) = C \int_{-\infty}^{+\infty} d
\lambda \exp \left[\sum_{j=2}^{\infty} \frac{i^j \lambda^j}{j!}
\frac{V}{(2 \pi)^3} \int d^3 \! p m_j(p) \epsilon_{p}^j
\right] ~.\label{tWmce0}
\end{equation}
It is convenient to introduce the notation
\begin{equation}
\sigma^2 = \frac{V}{(2 \pi)^3} \int d^3 \! p m_2(p)
\epsilon_{p}^2~.
\end{equation}
After replacing the integration variable,
$\lambda = x/\sigma$,
the integral (\ref{tWmce0}) takes the form
\begin{equation}\label{intx}
\tilde{W}_{m.c.e.}(0) = \frac{C}{\sigma} \int_{-\infty}^{+\infty}
d x \exp \left(- \frac{x^2}{2} \right) \exp \left[
\sum_{j=3}^{\infty} \frac{i^j \kappa_{j j}}{j!} x^j \right]~,
\end{equation}
where
\begin{equation}
\kappa_{l j} = \frac{1}{\sigma^j} \frac{V}{(2 \pi)^3} \int
d^3 p\;  m_l(p)\; \epsilon_{p}^j~.
\end{equation}

It is easy to see that
\begin{equation}
\kappa_{l j} \propto V^{1-j/2},
\end{equation}
i.e. the coefficients $\kappa_{j j}$, $j \ge 3$ ,in the second
exponent of (\ref{intx}) become small at $V \rightarrow \infty$.
We expand the second exponential function in Eq.~(\ref{intx}) and
perform the integration:
\begin{eqnarray}
\tilde{W}_{m.c.e.}(0) &=& \frac{\sqrt{2 \pi}  C}{\sigma} \left[ 1
+ \left( \frac{1}{8} \kappa_{4,4} - \frac{5}{24} \kappa_{3,3}^2
\right) + O(V^{-2}) \right] ~.
\end{eqnarray}
Similarly,
\begin{eqnarray}
\left. \frac{d \tilde{W}_{m.c.e.}(\nu)}{d \nu} \right|_{\nu=0}
 &=&
- \frac{ i C}{\sigma} \int_{-\infty}^{+\infty} d x
\left(\overline{N} + \sum_{j=2}^{\infty} \frac{i^{j-1} \kappa_{j
(j-1)}}{(j-1)!} x^{j-1} \right) \times \exp \left(- \frac{x^2}{2}
\right) \exp \left[ \sum_{j=3}^{\infty} \frac{i^j \kappa_{j
j}}{j!}  x^j
\right] \nonumber \\
&=& \frac{i \overline{N} C \sqrt{2 \pi} }{\sigma} \left[ 1 + \left(
\frac{\kappa_{2,1} \kappa_{3,3}}{2 \overline{N}} -
\frac{\kappa_{3,2}}{2 \overline{N}} -\frac{5}{24} \kappa_{3,3}^2 +
\frac{1}{8} \kappa_{4,4} \right) +  O(V^{-2}) \right]
\end{eqnarray}
and
\begin{eqnarray}
\left. \frac{d^2 \tilde{W}_{m.c.e.}(\nu)}{d \nu^2} \right|_{\nu=0}
 &=&
- \frac{C}{\sigma} \int_{-\infty}^{+\infty} d x
\left[\sum_{j=2}^{\infty} \frac{i^{j-2} \kappa_{j (j-2)}}{(j-2)!}
x^{j-2} \;+\; \left(\overline{N} + \sum_{j=2}^{\infty}
\frac{i^{j-1} \kappa_{j (j-1)}}{(j-1)!} x^{j-1}
\right)^2 \right] \nonumber \\
& & \times \exp \left(- \frac{x^2}{2} \right) \exp \left[
\sum_{j=3}^{\infty} \frac{i^j \kappa_{j j}}{j!}  \, x^j
\right] = - \frac{\overline{N}^2 C \sqrt{2 \pi} }{\sigma}
\left[ 1 + \left(
\frac{\kappa_{2,0}}{\overline{N}^2} \right. \right. \nonumber \\
& & \left. \left. - ~\frac{\kappa_{2,1}^2}{\overline{N}^2} -
\frac{\kappa_{3,2}}{\overline{N}} + \frac{\kappa_{2,1}
\kappa_{3,3}}{\overline{N}} - \frac{5}{24} \kappa_{3,3}^2 +
\frac{1}{8} \kappa_{4,4} \right) +  O(V^{-2}) \right] ~.
\end{eqnarray}

Now the moments of the probability distribution $W_{m.c.e.}(N)$
can be calculated:
\begin{equation}\label{Nmce}
\langle N \rangle_{m.c.e.} = \frac{1}{i \tilde{W}_{~m.c.e~}(0)}
\left. \frac{d \tilde{W}_{m.c.e.}(\nu)}{d \nu} \right|_{\nu=0} =
\overline{N} \left[ 1 + \frac{1}{2 \overline{N}}\left(\kappa_{2,1}
\kappa_{3,3} - \kappa_{3,2} \right) + O(V^{-2}) \right]
\end{equation}
and
\begin{equation}\label{N2mce}
\langle N^2 \rangle_{m.c.e.} = - \frac{1}{\tilde{W}_{m.c.e.}(0)}
\left. \frac{d^2 \tilde{W}_{m.c.e.}(\nu)}{d \nu^2} \right|_{\nu=0}
= \overline{N}^2 \left[ 1 + \left( \frac{\kappa_{2,0} -
\kappa_{2,1}^2}{\overline{N}^2} + \frac{\kappa_{2,1} \kappa_{3,3}
- \kappa_{3,2}}{\overline{N}} \right) + O(V^{-2}) \right].
\end{equation}
As it should be, $\langle N \rangle_{m.c.e.}$ and $\langle N^2
\rangle_{m.c.e.}$ approach, respectively, the grand canonical
values $\overline{N}$ and $\overline{N}^2$ in the thermodynamical
limit. The leading finite-volume corrections decay as $1/V$. 
Calculation of higher-order corrections is straightforward.

As is seen,  the correction terms containing $\kappa_{2,1}
\kappa_{3,3} - \kappa_{3,2}$ contribute both to $\langle N
\rangle_{m.c.e.}$ and $\langle N^2 \rangle_{m.c.e.}$, but cancel
each other in the scaling variance,
\begin{equation}
\omega_{m.c.e.} = \frac{\langle N^2 \rangle_{m.c.e.} - \langle N
\rangle_{m.c.e.}^2} {\langle N \rangle_{m.c.e.}} =
\frac{\kappa_{2,0} - \kappa_{2,1}^2}{\overline{N}}
+ O(V^{-1}) = \frac{\int d^3p ~v_p^2 \int d^3p ~v_p^2
\epsilon_{p}^2 - \left( \int d^3p ~v_p^2 \epsilon_{p}
\right)^2}{\int d^3p ~v_p^2 \epsilon_{p}^2 \int d^3p ~n_p } +
O(V^{-1})~,
\end{equation}
so that the result for fluctuations is indeed the same as in the previous
section (\ref{omega-mce}).

\section{Summary}
   We have proposed a new method
   for a microcanonical treatment of quantum
   gases near a thermodynamic limit.
   The method is based on the analysis of moments of the particle number
distribution in the microcanonical ensemble. For particle number
fluctuations in the thermodynamic limit it leads to the same
results as the microscopic correlator method \cite{steph2}.
However, the new method is more mathematically rigorous and
consistent,  and it elucidates some subtleties.It gives,
therefore, a justification of our previous  findings
\cite{mce-fluc} that the scaled variance for particle number
fluctuations in the microcanonical ensemble is different from that
in the grand canonical ensemble even in the thermodynamic limit.
Along with fluctuations, the new method allows calculating
the finite-volume corrections to the thermodynamic quantities in
the microcanonical ensemble. This can not be done within the
microscopic correlator
 calculations. Our approach can be
straightforwardly extended to the system of
charged particles with exact
charge conservation laws taken into account.



 \begin{acknowledgments}
We would like to thank F.~Becattini, A.I.~Bugrij, J.~Cleymans,
T.~Cs\"org\H{o},
M.~Ga\'zdzicki, W.~Greiner, A.~Ker\"anen,
I.N.~Mishustin, St.~Mr\'owczy\'nski, K.~Redlich,
L.~Turko, L.M.~Satarov, E.V.~Shuryak,
Y.M.~Sinyukov, and H. St\"ocker for useful discussions and comments.
 We are grateful to M.~Stephanov for a
discussion that greatly influnced this work.
The work was partially supported by US Civilian Research
and Development Foundation (CRDF) Cooperative Grants Program,
Project Agreement UKP1-2613-KV-04.
\end{acknowledgments}

\appendix

\section{Fourier transform of a probability distribution}
Let us consider a probability distribution $W(x)$. The Fourier
 transform of this distribution is given by:
\begin{equation}\label{Wx}
\tilde{W}(y) = \int d x e^{i x y} W(x)~,
\end{equation}
or, for a discrete  variable $x$,
\begin{equation}\label{Wtx}
\tilde{W}(y) = \sum_x e^{i x y} W(x)~.
\end{equation}
It is easy to check that the derivatives of the function
$\tilde{W}(y)$ are related to the average values of $x^k$:
\begin{equation}
\langle x^k \rangle = \frac{1}{\tilde{W}(0)} \left. \left(
\frac{1}{i}  \frac{d}{d y} \right)^k  \tilde{W}(y) \right |_{y=0}~.
\end{equation}

The central moments of the distribution $W(x)$ can be conveniently
calculated using $\log \tilde{W}(y)$:
\begin{equation}
m_k =  \left. \left( \frac{1}{i}  \frac{d}{d y} \right)^k  \log
\tilde{W}(y) \right |_{y=0}~,
\end{equation}
where
\begin{eqnarray}
m_1 &=& \langle x \rangle ~,\\
m_2 &=& \langle (x -  \langle x \rangle)^2 \rangle~, \\
m_3 &=& \langle (x -  \langle x \rangle)^3 \rangle ~,\\
m_4 &=& \langle (x -  \langle x \rangle)^4 \rangle -
        3 \langle (x -  \langle x \rangle)^2 \rangle^2 ~,\\
& & \mbox{\hspace{-1.3cm} etc.}           \nonumber
\end{eqnarray}

\end{document}